\documentstyle[aps,multicol,epsfig]{revtex}

\newcommand{\e}{{\rm e}}
\newcommand{\dd} {{\rm d}}

\newcommand{\eps}{\varepsilon}

\begin{document}
\draft

\title{Delocalization in an open one-dimensional chain in an imaginary
vector potential}

\author{Igor V. Yurkevich and Igor V. Lerner}
\address{
School of Physics and Astronomy, University of
Birmingham, Birmingham~B15~2TT, United Kingdom
}\date{\today} \maketitle \begin{abstract}
We present first results for the transmittance, $T$,  
through a 1D disordered system with an imaginary vector potential, $ih$,
which provide a new analytical criterion for a delocalization transition
in the model. It turns out that the position of
the critical curve on the complex energy plane (i.e.\ the curve
where an exponential decay of
$\left<T\right>$  is changed by a power-law one)
is different from that obtained previously from the complex energy spectra.
Corresponding curves for $\left<T^n\right>$
or $\left<\ln T\right>$ are also different. This happens because of 
different scales of the exponential decay of one-particle Green's functions
(GF) defining the spectra and many-particle GF governing transport 
characteristics, and reflects higher-order correlations in localized
eigenstates of the non-Hermitian model.
\end{abstract}

\begin{multicols}{2}

Non-Hermitian models with disorder have attracted a great deal of
attention, e.g.\ in context of open quantum-mechanical (or optical)
systems connected to reservoirs\cite{VWZ,FPY,Zhang}
or random walks in disordered media\cite{ArNel},
or Gaussian ensembles of non-Hermitian random matrices\cite{Somm:88}.
The discovery by Hatano and Nelson \cite{HN} 
of a delocalization transition in a simple 
Anderson model with an imaginary
vector potential has led to a new burst of 
activity in this area\cite{ChalWang}
and also focused research interest on  this
particular model\cite{Efet:97a,BSB:97,Silv:98,HataNels} and related problems
\cite{MSA:98}.

The numerical results on the complex spectra of the model
\cite{HN} have been confirmed analytically, both
for the zero-$D$ \cite{Efet:97a} and $1D$  \cite{BSB:97} cases.
The fact that the transition into complex
spectra in this model
is indeed a delocalization transition has been convincingly
illustrated by direct numerical investigations \cite{HataNels}
of the product
of left and right eigenfunctions corresponding to the same 
{\it complex} eigenvalue. This product,  which should enter
into any physical observable, has shown to be delocalized
\cite{HataNels}, in spite of the fact that   
eigenfunctions by themselves still look localized even in complex
spectra \cite{Silv:98}. It demonstrates that 
the description of the 
localization-delocalization transition may depend on the choice
of the localization criterion applied, in contrast to the standard
Hermitian model. This makes it interesting to look at some analytical criterion
of the transition. 

One of the best ways to address this problem is to calculate 
the transmittance $T$ or, equivalently, the
Landauer conductance.
Short of a direct analytical calculation of the wavefunction distribution,
 an analytical dependence of $T$ on the system size $L$ provides the most
straightforward criterion of the delocalization transition.
It allows one to investigate how the localization length
$\xi$ depends on the particle energy and how it diverges in approaching
the transition. It is well known that the
transmittance through a 1D Hermitian disordered wire decays as
\begin{equation}
\label{T} T= a(L)\exp\left[-\frac{L}{\xi}\right]\,,\qquad
a(L)\propto L^{-3/2}\,,
\end{equation}
indicating that all the wavefunctions are localized with the localization
length $\xi$ (depending on energy).
The existence of the delocalization transition should be manifested
by the divergence in $\xi$ at the mobility edge. 

In this Letter, we present the analytical results for the transmittance
$T$ which confirm the existence of the delocalization transition
in the $1D$  model introduced in \cite{HN}.
However, properties of the transition and {\it even the position of the 
mobility edge} on the complex energy plane turn out to be 
different from those expected from the knowledge of the complex spectra
alone. Before describing this, note that $T$ can
only be analyzed for an {\it open} system. If one considered a
system with free boundaries, the 
imaginary vector potential, $h$,  could be 
wiped out by a gauge transformation. We consider a disordered
sample attached to ideal leads, with the logarithmic derivative of the 
wave function being continuous at the boundaries. Although $h$ can be
eliminated from the {\it sample} by the gauge transformation, 
the requirement that the wave functions remain finite in the leads imposes
a nontrivial constraint on the class of allowed functions,
similar to that imposed by the periodic boundary conditions.

Although the open geometry for
the model  with an
imaginary vector potential could not be directly related to the problem of
depinning of flux lines in a superconductor with columnar defects
\cite{NV:93}, which was the original motivation for this model \cite{HN},
the subsequent interest has arisen
due to a general character of the observed hyper-sensitivity of a disordered
system to the presence of even a small non-Hermiticity. 
Thus we believe that the analytical description in the case of
the open geometry can be very useful, even if not directly related
to the original model of depinning\cite{NV:93}.

The model is described by
the Schr\"{o}dinger equation
\begin{equation}
\label{H}
{\mathcal H}\psi(x) \equiv
\biggl[
-\left( \frac{\dd}{\dd x}-h\right)^{\!2} + \upsilon(x)\biggr]\!\psi(x)=
z\psi(x)\,.
\end{equation}
Here $\psi(x)\!\equiv\!\psi_{{\cal R}}(x)$
is the right eigenfunction of $\mathcal H$
corresponding to the complex energy $z$; in this model the left
eigenfunction at the same energy is
$\psi_{{\cal L}}(x;h)\!=\!\psi^*_{{\cal R}}(x;-h) $ \cite{HN}. 
The random potential, $\upsilon(x)$,
 is chosen to be Gaussian white-noise with zero mean
and variance $u^2$. It vanishes in 
the ideal leads attached to the sample at $x=0$ and $x\!=\!-L$. 
 The imaginary
vector potential, $ih$, is homogeneous inside the sample
and make take a different constant value (not necessarily $0$)
in the leads.  The logarithmic derivative of $\psi(x)$ is continuous
at the boundaries.


In the non-Hermitian case considered, we must find $T$ as a function of
complex energy $z$. Then the localization-delocalization transition
would reveal itself via the divergence of $\xi(z)$ at certain values
of $z$ (the mobility edge). Furthermore, for the values of $z$ at which
the wavefunctions are delocalized, the transmittance is expected
to have a non-exponential dependence on $L$. 

However,  solving the standard scattering problem for the Hamiltonian
(\ref{H}), we find that $T$ always remains exponential as in Eq.\ (\ref{T}),
albeit with $\xi $ increasing as a function of $h$. 
On the face of it, this result is quite surprising. It means that, were one
able to have an experimental realization of a 1D wire in the presence of
the imaginary vector potential, measuring a current through it would not
reveal an insulator-to-metal transition, at least as a function of the wire
length. The reason for this is rather simple: the energies of 
incident plane waves on the complex $z$ plane do not overlap with the energies
of the delocalized states. Although these states exist, they make
only an exponentially small contribution to the transmittance. 
Therefore, we need to generalize the scattering technique to be able to
detect the delocalized states. 

We will define the scattering amplitudes, $t$ and $r$, via the 
asymptotics of the wave function:
\begin{equation}
\label{ash}
\psi(x)=\cases{
t\e^{ik_+x}, & $x>0$\,;\cr
\e^{ik_+(x+L)} + r\e^{-ik_-(x+L)}, & $x<-L$\,.
}
\end{equation}
We suppose, for a moment, that the imaginary vector potential $h$
is the same both inside and outside the sample. Then the generalized 
wave-vector, $k_\pm$, is defined by
$$
k_\pm=\sqrt z \mp ih\equiv k\pm i(\kappa-h)\,,
$$
with $k$ and $\kappa$ being the real and imaginary parts of $\sqrt z$
(a branch with $\kappa>0$ is chosen). 
Both the incident and transmitted waves are not divergent at $x\to\infty$
only provided that $k_+ $ is real, i.e.\ on
 the curve
$S_0$ in the $z$-plane,
\begin{equation}
S_0: \quad \kappa = h\quad\Longleftrightarrow\quad
\Re{\rm e } \, z = \left(\Im{\rm m }\, z/2h\right)^2 - h^2\,,
\label{S0}
\end{equation}
which happens to be also the DoS support curve in the absence of the 
impurity potential $\upsilon$. The DoS support inside the sample 
($\upsilon\ne0$) is, however,  entirely different.

To illustrate this, we first consider a simple case of the one-particle 
 Green function (GF), $G(z)$,  which is the GF of the 
Schr\"{o}dinger equation (\ref{H}).
A straightforward calculation in the absence of the disorder,
$\upsilon(x)=0$, yields 
\begin{equation}
\label{GF0}
G_0(x,0;z)\!=\!\frac{1}{2i\sqrt{z}}\!\times\!\!
\cases{
\theta(\kappa\!-\!h\!)\e^{ik_+x}&$x\ge0$ \cr
\e^{-ik_-x} -\theta(\!h\!-\!\kappa\!)\e^{ik_+x}
&$x\le0$
}
\end{equation}
As a function of $z$, $G_0$ is discontinuous
at $\kappa = h$ which means that all the eigenvalues lie on the curve
$S_0$ in the $z$-plane defined by Eq.\ (\ref{S0}).
This can easily be seen from the well known formula for the density
of states (DoS), $\nu(z)=\pi^{-1}\partial_{z^{\!*}}G(0,0;z)$. Thus, 
the DoS support lies on the parabola along the $\Re{\rm e } \, z$
axis on the $z$ plane.

 In the presence of the disorder  it
is straightforward to show that in the quasi-classical regime, 
$k\gg \kappa, \ell^{-1}, h$, the ensemble-averaged GF of
Eq.\ (\ref{H}) remains of the same form as $G_0$, Eq.\ (\ref{GF0}), with 
the only change
\begin{equation}
\label{GF}
 k_\pm\to  k_\pm \pm i/2\ell(k)\,,
\end{equation}
the same as for the Hermitian problem, with $\ell\equiv\ell(k)=2k^2/u^2$
being a mean free path.
The  discontinuity line of the GF defines the
DoS support inside the sample:
\begin{equation}
\label{supp}
S_1: \quad \kappa = \mbox{max}\{h-1/2\ell(k),\,0\}\,.
\end{equation}
The DoS support lines inside and outside the
sample (with the same imaginary vector potential, $ih$, in both 
regions), Eqs.\ (\ref{supp}) and (\ref{S0}), 
 do not overlap at any disorder. If we put $h=0$ outside the sample,
then all states are on the line 
$\kappa=0$ which only intersects $S_1$ at a single point.
In order to detect delocalized
 states inside the sample,
 $k_+$ in the incident wave, Eq.\ (\ref{ash}) should be allowed to take on
 any complex value.
To tune $k_+$, one could formally apply the imaginary potential $ih_0$ outside
of the sample. Then, by changing $h_0$, one
would shift the DoS support line (\ref{S0}) 
in such a way that it will intersect with $S_1$ (see Fig.~1). Note that any
physical quantity, like conductance $\propto T$, would be defined
by the ratio of the wave functions on the sample boundaries, and thus
will be totally unaffected by $h_o$.

One would expect that $S_1 $ also defines scattering properties of 
the disordered system, in particular the transmittance $T$ and
thus the mobility edge where $\xi(k, \kappa)\to\infty$. Indeed, Eq.\ 
(\ref{supp}) defines a parabola on the $z$ plane which is an exact counter-part
of the bubble-like curve analytically found in \cite{BSB:97} for the lattice
variant of the model. 
However, the situation turns out to be not that simple. What
we have found is that 
the transmittance cumulants, $\left<T^n\right>$, are characterized
by the whole set of the critical curves, $C_n$ (see Fig.~1). 

\begin{figure}
\epsfxsize=.9\hsize
\epsffile{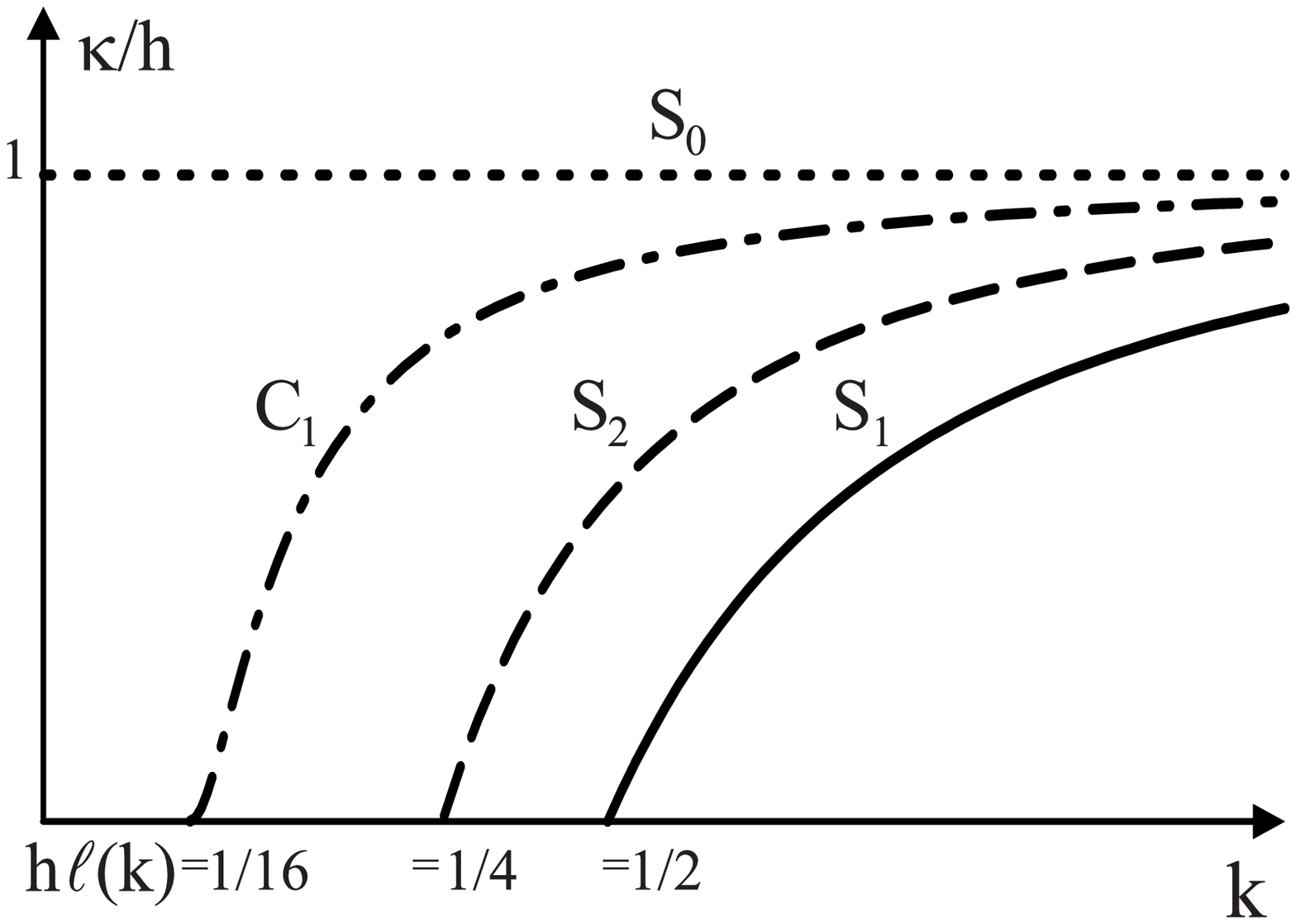}

{\small \setlength{\baselineskip}{10pt} FIG.\ 1.
 Phase diagram in $\sqrt{z}$-plane:
 the  DoS support lines for a pure system, $S_0$,
and a disordered system, $S_1$;
the critical curves for $\left<\ln T\right>$, $S_2$, 
and $\left<T\right>$, $C_1$; $\ell(k)=2k^2/u^2$, Eq.~(\ref{GF}).}
\end{figure}


This multitude of the critical curves is due to the
absence of  self-averaging in 1D\cite{Mel:81}.
It is  $\ln T$ 
which is normally-distributed in the Hermitian case.
The lognormal distribution of $T$ leads to all the moments
$T^n$ to have the same decay exponent $-L/\xi$ (which is not scaled
with $n$), where $\xi$ is the localization length.
 It is the two-particle GF
which decays $\propto\e^{-L/\xi}$ and thus contains the information
about the localization, while the decay of the one-particle GF,
Eq.\ (\ref{GF}), is 
only due to the  loss of the momentum direction in
elastic scattering, which is the same in any dimensionality. The 
self-averaging $\ln T$ is governed by its own exponent
different from that in the  one-particle GF by substituting
the transport scattering
time, $\ell_{\text{tr}}=2\ell$, for $\ell$.
In the presence of $ih$, the exponential decay is replaced
by $\e^{(nh-1/\xi)L}$ for $T^n$, or by $\e^{(h-1/2\ell)L}$
for the one-particle
GF and  ${(h\!-\!1/2\ell_{\text{tr}})L}$ for $\ln T$.
Then for $h\!>\!1/\xi$ (or $1/2\ell,\:\;1/2\ell_{\text{tr}}$, respectively)
 no small readjustment of localized states could
prevent a mismatching of boundary conditions for a closed system (or 
an exponential increase in one of the leads for an open system). 
This defines a set of critical curves, Fig.~1,
 which are different for all these quantities.

Now we outline our procedure of rigorous analytical deriving of these
results. By introducing vector $\Psi$, 
\begin{equation}
\label{5}
\Psi=\left(\begin{array}{c}\psi_+\\ \psi_-\end{array}\right),
\quad
\psi_{\pm}\equiv
\pm \frac{1}{2i\sqrt{z}}
\frac{\dd\psi}{\dd x}+ik_{\mp}\psi\,,
\end{equation}
we reduce the Schr\"{o}dinger equation (\ref{H})
to the first-order matrix equation:
\begin{equation}
\label{6}
\frac{\dd\Psi}{\dd x}=\mu \Psi, \quad
\mu(x)\equiv
h+i\sqrt{z}\sigma_3 +\frac{\upsilon(x)}{2i\sqrt{z}}(\sigma_3 +
i\sigma_2) \,,
\end{equation}
where $\sigma_i$ are the Pauli matrices. The corresponding
transfer-matrix, $m$,  is defined by the relation
\begin{equation}
\label{TM}
\Psi(x)= m(x,x') \Psi (x')
\end{equation}
and obeys the following first order differential equation
\begin{equation}
\label{TME}
\frac{\dd m(x,x')}{\dd x}=\mu (x) \cdot m(x,x')\,,
\end{equation}
which is subject to the initial condition $m(x,x)=1$.

Taking into account that the values of $\Psi$ at the boundaries,
$\Psi^T(-L)=(1,r)$ and $\Psi^T(0) = (t,0)$
are related via Eq.\ (\ref{TM}) 
by $m(-L,0)$, we express
the reflection and transmission amplitudes in Eq.\ (\ref{ash})
as follows
\begin{equation}
\label{rel}
t=\frac{1}{m_{11}\left(-L,0\right)}, \quad
r=\frac{m_{21}\left(-L,0\right)}{m_{11}\left(-L,0\right)}\,.
\end{equation}
From Eq.(\ref{TME}) we can extract two coupled equations for the scattering
amplitudes $r$ and $t$:
\begin{eqnarray}
\frac{\dd r}{\dd L}&
=&2i\sqrt{z}\,r+\frac{\upsilon (-L)}{2i\sqrt{z}}
\left(1+r\right)^2, \nonumber\\[-5pt]\label{ie}\\[-5pt]\nonumber
\frac{\dd t}{\dd L}
&=&
(h+i\sqrt{z})t+\frac{\upsilon (-L)}{2i\sqrt{z}}t
\left( 1 + r \right),
\end{eqnarray}
which obey the boundary conditions at the right end of the sample,
$r(0)=0$ and $t(0)=1$. By performing the ensemble averaging, one
can easily derive equations for the mixed moments $\langle
T^nR^{n'}\rangle$. As $T+R\ne1$ in the 
non-Hermitian case, $R\equiv |r|^2$
cannot be simply excluded. For our purposes, it is sufficient to 
consider only conditional averages, $P_n(L)=
\langle T^n(L,R)\delta\left(R(L)-R\right)\rangle$ so that the moments
$\langle T^n\rangle$ could be found by integrating over all $R$. 
As the reflectance $R$ is an auxiliary quantity, it is convenient
to introduce a new variable, $\chi$,  by $R=\tanh^2(\chi/2)$. In the
absence of the imaginary potential, $L/\chi$ is equal to the localization
length\cite{Bee:93a}.
The conditional probability $P_n(\tau, \chi)$ in new variables (with
$\tau\equiv L/2\ell$) obeys the following Fokker-Planck equation,
\begin{equation}
\label{FP}
\frac{\partial P_n}{\partial \tau}=\frac{\partial}{\partial \chi}
\left[
\frac{\partial}{\partial \chi }+\frac{\partial \Omega_n}{\partial \chi}
\right]P_n + V_n P_n
\,,
\end{equation}
where the initial condition is $P_n(\tau\!=\!0)=\delta(\chi)$, and
\begin{eqnarray}
V_n(\chi)&\equiv&n^2\tanh^2\frac{\chi}{2} -n - 4n(\kappa - h)\ell\,,
 \nonumber\\[-8pt]\label{omega}\\[-8pt]\nonumber
\Omega_n(\chi)&\equiv&4\kappa \ell  \cosh \chi 
-\ln\sinh \chi + 4n\ln\cosh\frac{\chi}{2}
\,,
\end{eqnarray}
which is derived from Eqs.\ (\ref{ie}) for $k\gg\kappa, h, \ell^{-1}$
in a way similar to that for 
the imaginary scalar potential \cite{FPY}. The moments of $T$ can
be found by the integration: 
\begin{equation}
\label{mom}
\langle T^n(\tau)\rangle=\int_0^{\infty}\dd \chi P_n(\chi,\tau)
\,.
\end{equation}
As $\ln T$ is well known to be a self-averaging quantity in 1D, 
the Lyapunov exponent,
\begin{equation}
\label{Lyap}
\lambda =-\lim_{L\rightarrow\infty}\frac{\langle\ln T\rangle}{L},\quad
\langle\ln T\rangle=\lim_{n\rightarrow 0}\frac{\ln\langle T^n \rangle }{n}\,,
\end{equation}
found via the standard `replica' trick, gives the `best' representation of the
inverse localization length.

Equation (\ref{FP}) cannot be solved exactly and we map
it onto the imaginary-time
Schr\"{o}dinger equation:
\begin{equation}
\label{Sch}
-\frac{\partial \Phi_n}{\partial \tau} =-\frac{\partial^2\Phi_n}{\partial 
\chi^2} +U_n(\chi)\Phi_n\,.
\end{equation}
Here $\Phi_n(\chi,\tau)=P_n(\chi,\tau)\exp[\Omega_n(\chi)]$, and the effective
potential 
$U_n\equiv(\Omega'_n)^2/4-\Omega''_n/2-V_n$  reduces to
\begin{eqnarray}
\label{pot}
\nonumber
U_n(\chi)&=&\frac{1}{4}- \frac{1}{4\sinh^2\chi} - 4nh\ell\\
&+&{4(\kappa\ell)^2}\sinh^2\chi + 4\kappa\ell (n-1)\cosh \chi 
\,.
\end{eqnarray}
Although the eigenfunctions $\Phi_n^j(\chi)$ of Eq.\ (\ref{Sch}), where $
\Phi_n(\chi,\tau)\equiv \sum_j\Phi_n^j(\chi)\,\e^{-\eps_n^j\tau}$, cannot be
found exactly, the form of the effective potential, Eq.\ (\ref{pot}), makes
possible to find the long-$\tau$ limit of $ \Phi_n(\chi,\tau)$, and thus of
$P_n(\chi, \tau)$. A sharp increase in $U_n$ at $\chi\agt\ln(1/\kappa\ell)$
makes the spectrum of Eq.\ (\ref{Sch}) discrete, with a gap of order $1$
separating the ground and excited states. Therefore, only the lowest
eigenstates contribute to $ \Phi_n(\chi,\tau)$ and thus to $\langle
T^n\rangle$ for $\tau\equiv L/2\ell\gg1$. It is easy to verify that for $n=0$
the ground state has the energy $\eps_o^o=0$ and the eigenfunction
\begin{equation}
\label{GS}
\displaystyle \Phi_0^{0}(\chi)=\frac{\e^{-\frac{1}{2}\Omega_0(\chi)}}
{\left[\int \dd \chi e^{-\Omega_0(\chi)}\right]^{1/2}}.
\end{equation}
For $n\ll1$, by treating the $n$-dependent part of the potential (\ref{pot}) as
a perturbation, one finds $\Psi_n^{0}\approx\Psi_0^{0} $ and
$\eps^0_n=n[1+4(\kappa-h)\ell]$. This leads, via Eq.\ (\ref{Lyap}), to exact
\begin{eqnarray}
\label{Lyap2}
\lambda &=& - \frac1{2\ell}\lim_{\tau\rightarrow\infty}\lim_{n\rightarrow 0}
\frac{\eps_n^0}{n}= 2(\kappa-h)+ \frac1{2\ell}\,.
\end{eqnarray}
The Lyapunov exponent $\lambda(z)$ vanishes (i.e.\ the localization length
diverges) on the curve (see Fig.~1)
\begin{equation}
S_2: \quad \kappa = h - {1}/{4\ell(k)}\,.
\end{equation}
This curve, as expected, could be obtained from $S_1$,
 Eq.(\ref{supp}), by substituting there 
 $\ell_{\text{tr}}=2\ell$ for $\ell$.

In contrast to $\langle \ln T\rangle$, the moments $\langle T^n\rangle$
cannot be found exactly for $n\ge1$. However, as $\kappa\ell\ll1$,
approximate eigenfunctions can be found by separating 
$\chi\alt1$ and $\chi\agt\ln(\kappa\ell)^{-1}$ scales and matching 
appropriate solutions. Thus we find the
moments with $n\ll(\kappa\ell)^{-1}$:
\begin{equation}
\label{small}
\left<T^n(\tau;z)\right> = c_n
\left({1}/{\tau}+{1}/{\tau_c}\right)^{3/2}
\e^{-\left[\frac{1}{4}-2nh\ell+\frac{1}{\tau_c}\right]\tau},
\end{equation}
$
\mbox{where }
\tau_c(n)\!=\!
\pi^{-2}\left[\ln(1/2\kappa\ell) -\psi(n\!-\!{1}/{2})\right]^2,
$ $\psi$ is the di\-gam\-ma function, 
$c_n=\left[(2n\!-\!3)!!\right]^2\!\pi^{5/2}2^{1\!-\!2n}\left[(n\!-\!1)!\right]^{-1}$.
For moments with $n\gg(\kappa\ell)^{-1}$, the preexponential factor 
is proportional to $\sqrt{\kappa\ell/n}$, while the exponent becomes
$1/4-2n\ell(h-\kappa)$.
One finds from Eq.\ (\ref{small})
that the $n$-th exponent vanishes on the curve $C_n$ 
(which is drawn for $n=1$ in Fig.~1 and is qualitatively the same for 
any $n$):
\begin{equation}
C_n: \quad \kappa = \frac{1}{2n\ell (k)}\e^{-\left[2nh\ell
(k)-1/4\right]^{-1/2}}.
\end{equation}
The  corresponding transmittance moment on this curve 
close to the point where
$C_n$ hits the $k$-axis on the $k$--$\kappa$ plane, i.e.\ for 
 $h\ell (k)-1/8n \ll 1$, is given by
\begin{equation}
T_n(\tau;k) = c_n \left(\tau^{-1}+2nh\ell (k)-1/4\right)^{3/2}
\end{equation}
Such a power-low behavior, $\left<T^n\right>\propto
 (\ell/L)^{3/2}$, 
is typical for a critical regime at the metal-insulator transition point.
What is unusual is that each moment reaches the criticality at a different
point in the $z$ plane.

The existence of a set of different critical curves means that the 
delocalization transition, say in $\left<T\right>$, happens in the point
in the $z$ plane where the DoS support for a corresponding closed system, 
$S_1$, is still on the real axis (Fig.~1)
 and therefore all products of the left
and right eigenfunctions are still localized. 
This manifests the existence of some higher-order correlations between
the localized states, similar to the anti-correlations between the amplitudes
of $\psi_{\cal R}(z)$ and $\psi_{\cal L}(z)$ which ensures the delocalization
of their product on the curve $S_1$, despite each of these function by
itself is still `localized', i.e.\ ooccupies only a very small part of the
sample\cite{HataNels}. As the properties of the localized states should not
be strongly dependent on boundary conditions, these 
 higher-order correlations between them could also exist for the closed
disordered systems with the imaginary vector potential.

\acknowledgments
This work has been supported by the 
EPSRC grant GR/K95505.

 \end{multicols}


\vspace*{-.5cm}

\begin{references}
\vspace*{-1cm}

\bibitem{VWZ}
J.~J.~M. Verbaarschot, H.~A. Weidenm\"{u}ller, and M.~R. Zirnbauer, Phys. Rep.
  {\bf 129}, 367 (1985);
S.~Iida, H.~A. Weidenm\"{u}ller, and J.~Zuk, Phys. Rev. Lett. {\bf 64}, 583
  (1990).
 
\bibitem{FPY}
V.~Freilikher, M.~Pustilnik, and I.~Yurkevich, Phys. Rev. Lett. {\bf 73}, 810
  (1994);
Phys. Rev. {\rm B} {\bf 50},
  6017 (1994).
 
\bibitem{Zhang}
Z.~Q. Zhang, Phys. Rev. {\rm B} {\bf 52}, 7960 (1995).
J.~C.~J. Paasschens, T.~S. Misirpashaev, and C.~W.~J. Beenakker, {\it ibid},
 {\bf 54}, 11887 (1996).
 
\bibitem{ArNel}
J.~A. Aronovitz and D.~R. Nelson, Phys. Rev. {\rm A} {\bf 30}, 1948 (1984);
D.~S. Fisher {\it et al}, Phys. Rev.
  {\rm A} {\bf 31}, 3841 (1985);
V.~E. Kravtsov, I.~V. Lerner, and V.~I. Yudson, J. Phys. {\rm A} {\bf 18}, L703
  (1985), {Sov. Phys. JETP} {\bf 64}, 336 (1986);
J.~P. Bouchaud and A.~Georges, Phys. Rep. {\bf 195}, 127 (1990).
 
\bibitem{Somm:88}
H.~J.~Sommers {\it et al}, Phys. Rev. Lett. {\bf
  60}, 1895 (1988);
F.~Haake {\it et al}, Z. Phys. {\rm B}
  {\bf 88}, 359 (1992);
S.~Hikami and A.~Zee, Nucl. Phys. {\rm B} {\bf 446}, 337 (1995);
J.~Feinberg and A.~Zee, {\it ibid}, {\bf 504}, 579 (1997).
 
\bibitem{HN}
N.~Hatano and D.~R. Nelson, Phys. Rev. Lett. {\bf 77}, 570 (1996).
 
\bibitem{ChalWang}
J.~T. Chalker and Z.~J. Wang, Phys. Rev. Lett. {\bf 79}, 1797 (1997);
Y.~V. Fyodorov, B.~A. Khoruzhenko, and H.~J. Sommers, {\it ibid}, {\bf 79}, 557,
  (1997);
N.~M. Shnerb and D.~R. Nelson, {\it ibid}, {\bf 80}, 5172 (1998);
C.~M. Bender and S.~Boettcher, {\it ibid}, {\bf 80}, 5243 (1998),
J.~T. Chalker and B.~Mehlig, {\it ibid}, {\bf 81}, 3367 (1998);
C.~Mudry {\it et al}, Phys. Rev. {\rm B}
  {\bf 58}, 13539 (1998).
 
\bibitem{Efet:97a}
K.~B.~Efetov, Phys. Rev. Lett. {\bf 79}, 491 (1997);
Phys. Rev. {\rm B} {\bf 56}, 9630 (1997).
 
\bibitem{BSB:97}
P.~W. Brouwer, P.~G. Silvestrov, and C.~W.~J. Beenakker, Phys. Rev. {\rm B}
  {\bf 56}, R4333 (1997);
I.~Y. Goldsheid and B.~A. Khoruzhenko, Phys. Rev. Lett. {\bf 80}, 2897 (1998).
 
\bibitem{Silv:98}
P.~G. Silvestrov, Phys. Rev. {\rm B} {\bf 58}, R10111 (1998),
\bibitem{HataNels}
N.~Hatano and D.~R.~Nelson, {\it ibid} {\bf 58}, 8384 (1998).
 
 
\bibitem{MSA:98}
C.~Mudry, B.~D.~Simons, and A.~Altland, Phys. Rev. Lett. {\bf 80}, 4257 (1998);
P.~W.~Brouwer {\it et al}, {\it ibid}, {\bf 81}, 862
  (1998).
 
\bibitem{NV:93}
D.~R. Nelson and V.~Vinokur, Phys. Rev. {\rm B} {\bf 48}, 13060 (1993).
 
\bibitem{Mel:81}
V.~I. Mel'nikov, { Sov. Phys. Sol. State} {\bf 23}, 444 (1981);
A.~A. Abrikosov, Sol. State Commun. {\bf 37}, 997 (1981).
 
 
\bibitem{Bee:93a}
C.~W.~J. Beenakker and B.~Rejaei, Phys. Rev. Lett. {\bf 71}, 3689 (1993);
Phys. Rev. {\rm B} {\bf 49}, 7499 (1994).
 

\end{references}
\end{document}